%% file: geofaham.tex
\documentclass[
  journal=medium,
  manuscript=technical-report,
  year=2026,
  volume=0,
]{cup-journal}
\usepackage{amsmath}
\usepackage[nopatch]{microtype}
\usepackage{booktabs}
\usepackage{amssymb}

\usepackage[T1]{fontenc}    
\usepackage{hyperref}       
\usepackage{url}            
\usepackage{amsfonts}       
\usepackage{nicefrac}       
\usepackage{graphicx}       
\usepackage{tikz}          
\usetikzlibrary{positioning,arrows.meta,fit,backgrounds}
\usepackage{standalone}    
\usepackage{makecell}

\usepackage{subcaption}
\usepackage{listings}
\usepackage{tcolorbox}
\tcbuselibrary{skins,breakable,listings}

\colorlet{tbrowcolor}{white}
\colorlet{tbheadcolor}{white}

\title{GeoFaham: Domain-Specialized Multi-Agent Geospatial Analytics for Disaster Response}

\author{Muhammad Sohail Danish\textsuperscript{\S}}
\affiliation{Microsoft AI for Good Lab}
\affiliation{Mohamed bin Zayed University of Artificial Intelligence}

\author{Gilles Q. Hacheme}
\affiliation{Microsoft AI for Good Lab}

\author{Akram Zaytar}
\affiliation{Microsoft AI for Good Lab}

\author{Satej S. Soman\textsuperscript{\S}}
\affiliation{Microsoft AI for Good Lab}

\author{Inbal Becker-Reshef}
\affiliation{Microsoft AI for Good Lab}

\author{Juan Lavista Ferres}
\affiliation{Microsoft AI for Good Lab}

\author{Salman Khan}
\affiliation{Mohamed bin Zayed University of Artificial Intelligence}

\author{Girmaw A. Tadesse}
\affiliation{Microsoft AI for Good Lab}
\email[Author8]{gtadesse@microsoft.com}


\makeatletter
\patchcmd{\@maketitle}
  {{\volumefont\cup@vol}, }
  {}
  {}{}
\renewcommand\paragraph{\@startsection{paragraph}{4}{\z@}%
  {3.25ex \@plus1ex \@minus.2ex}%
  {-1em}%
  {\sffamily\bfseries}}
\pretocmd{\cup@contact@details}%
  {{\textsuperscript{\S}Work done while the author was an AI Resident at the Microsoft AI for Good Lab,
  Nairobi, Kenya.}\par}%
  {}{}
\makeatother

\begin{document}
\maketitle

\input{sections/abstract}

\input{sections/introduction}
\input{sections/related}
\input{sections/architecture}
\input{sections/benchmark}
\input{sections/experiments}
\input{sections/conclusion}

\printbibliography

\clearpage
\appendix

\input{sections/appendix}

\input{sections/appendix_ui}
\end{document}

%% file: sections/abstract.tex
\begin{abstract}
Real-world geospatial decision support requires more than single-step retrieval or code generation. Disaster-response questions often involve multi-source workflows that combine spatial databases, map-derived features, satellite imagery, raster products, and reusable geospatial artifacts, where correctness depends on preserving spatial scope, temporal context, and intermediate execution state. While large language models and coding agents offer a promising interface for such workflows, they struggle with geospatial-specific failure modes, including handling large artifacts, selecting the wrong evidence source, losing spatial or temporal metadata, and applying invalid geometric operations. We introduce GeoFaham, a domain-specialized multi-agent system for geospatial disaster analytics. GeoFaham decomposes workflows into specialist agents for vector operations, map feature retrieval, imagery discovery, and raster analysis, coordinated by an Orchestrator with a dual-ledger control loop for dependency-aware planning, progress tracking, validation, and replanning. To support artifact-centric collaboration, GeoFaham uses a Response Store that persists full geospatial outputs and typed metadata outside the LLM context window while agents exchange compact identifiers and summaries. We also introduce GeoFaham-Bench, a 40-task benchmark grounded in 17 real-world disaster records, designed to evaluate multi-step geospatial reasoning, cross-source integration, raster-vector computation, artifact generation, and recovery under non-rigid ground truth. 
We evaluate GeoFaham on GeoFaham-Bench against strong general-purpose coding-agent baselines, including Codex, Claude Code, and OpenCode. GeoFaham achieves the best overall performance, scoring 0.940 versus the next-best 0.879, while using 7.5 times fewer tokens.
Our code and benchmark are publicly available
at \href{https://github.com/microsoft/GeoFaham}{https://github.com/microsoft/GeoFaham}.
\end{abstract}

%% file: sections/introduction.tex
\section{Introduction}
\label{sec:introduction}

Geospatial decision-making during disasters often requires executing complex, multi-source workflows rather than retrieving a single fact or generating an isolated map. This challenge is especially pronounced for policymakers and emergency-response teams in the Global South, where access to technical expertise, computational resources, and integrated geospatial infrastructure can be limited \citep{chen2021building,su151310624}. Answering an operational question may involve identifying the correct disaster extent, retrieving relevant infrastructure and road networks, selecting satellite imagery from appropriate pre- and post-event periods, performing spatial and raster operations, and producing reusable outputs such as maps, tables, GeoJSON layers, or GeoTIFFs. 
For example, determining whether hospitals near a flood remain potentially accessible requires combining hazard boundaries, healthcare facilities, and transportation networks while preserving valid spatial reference and distance calculations. Although GIS experts can manage these dependencies explicitly, natural-language systems must coordinate them automatically. A failure at any step such as selecting the wrong spatial extent, time window, or coordinate system can produce a plausible but operationally incorrect result.

\begin{figure}[t]
    \centering
    \includegraphics[width=.9\linewidth]{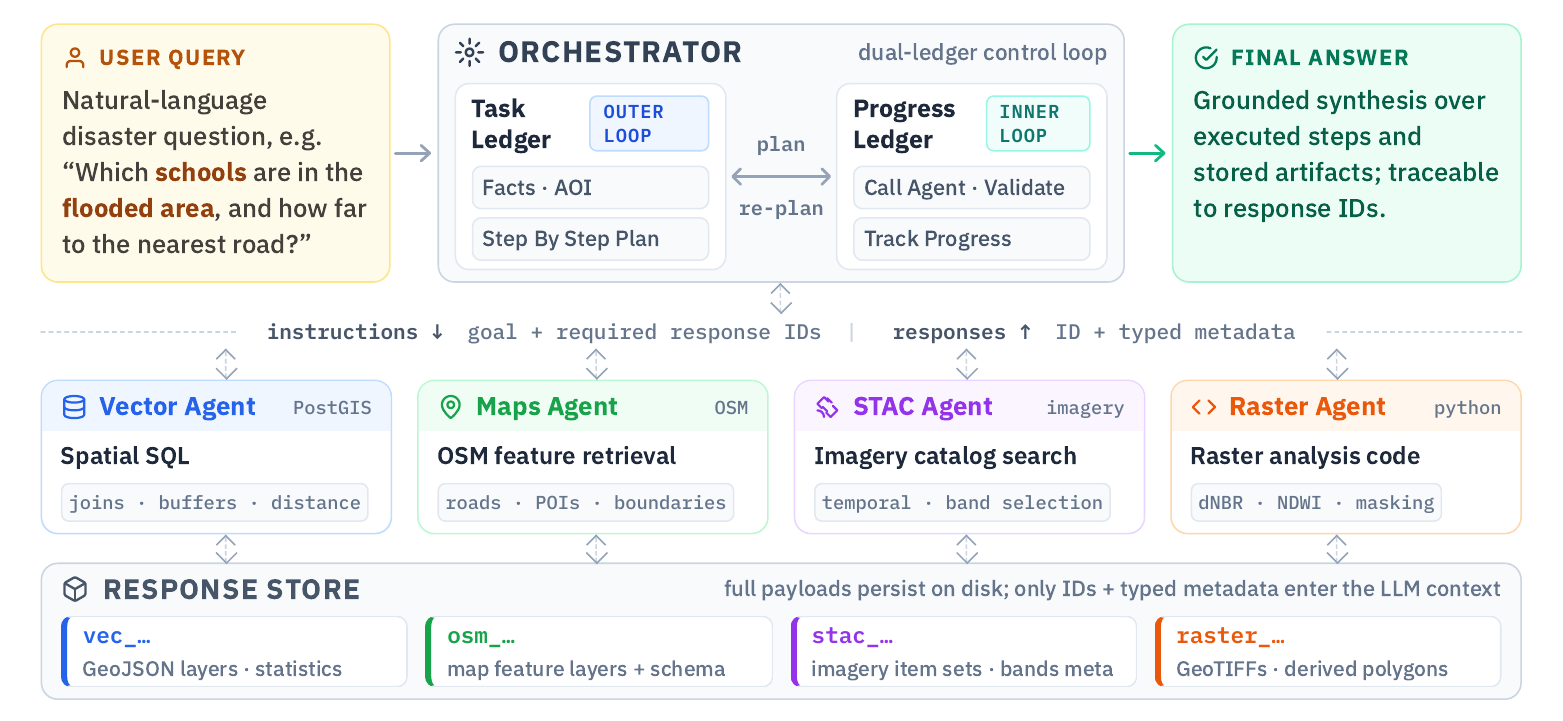}
    \caption{%
        GeoFaham overview. An Orchestrator coordinates four
        domain-specialized agents (Vector/PostGIS, Maps/OSM, STAC imagery
        retrieval, and Raster code generation) using nested control loops.
        The outer loop produces a task ledger and dependency-aware plan,
        while the inner loop tracks execution in a progress ledger with
        validation and replanning. Agents persist large geospatial outputs
        to a Response Store and exchange only compact identifiers and typed
        metadata through the Orchestrator.
    }
    \label{fig:geofaham-architecture}
\end{figure}

Recent advances in large language models (LLMs) have enabled agentic systems that can autonomously plan, invoke tools, and adapt to intermediate observations~\citep{wang2023survey,xi2023rise}. While such systems have shown strong performance on text-centric multi-step tasks, applying them to geospatial analytics introduces three key challenges. \textbf{(i) Domain complexity:} geospatial reasoning requires specialized knowledge of spatial databases and coordinate systems, satellite imagery and spectral information, map-derived data, and raster analysis. \textbf{(ii) Data heterogeneity:} operational pipelines combine vector layers, multi-band raster imagery, tabular summaries, and temporal products, each requiring distinct tools, formats, and validation procedures. \textbf{(iii) Scale and context:} geospatial artifacts can be large; GeoJSON layers may contain thousands of features, while raster products can reach gigabytes. Passing complete intermediate results through LLM context windows is therefore impractical, even though downstream correctness depends on preserving metadata such as coordinate reference systems, band names, acquisition timestamps, and spatial bounds.

General-purpose coding agents provide a promising execution framework for addressing these challenges. By decomposing tasks, generating and executing code, inspecting outputs, and revising plans based on intermediate observations, they offer a natural foundation for geospatial decision support. A straightforward approach is therefore to equip such an agent with geospatial libraries, spatial databases, map APIs, and satellite imagery catalogs, and ask it to solve geospatial workflows end-to-end. However, disaster-response workflows introduce failure modes that are not captured by successful code execution alone. An answer may appear fluent and its code may execute successfully, while the underlying analysis remains incorrect because the agent selected the wrong disaster footprint, used an inappropriate spatial boundary, searched imagery outside the relevant event window, computed distances in an invalid CRS, or relied on an intermediate artifact that was planned but never produced.

To address these challenges, we introduce GeoFaham (Figure~\ref{fig:geofaham-architecture}), a domain-specialized multi-agent system for geospatial disaster analytics. Rather than relying on a single generalist agent with broad tool access, GeoFaham decomposes geospatial workflows into four specialized roles defined by data source and operation type: structured spatial data querying, map feature retrieval, satellite imagery discovery, and raster analysis. A central Orchestrator coordinates these agents through a dual-ledger control loop. The outer loop extracts structured task facts and constructs a dependency-aware plan, while the inner loop tracks execution progress, verifies that required artifacts have been produced, and replans when execution stalls or a dependency cannot be satisfied.

A key feature of GeoFaham is artifact-aware collaboration. Agents do not pass large coordinate arrays, raster data, or complete imagery payloads through the LLM context. Instead, full outputs are persisted in a working memory called the Response Store, while the Orchestrator exchanges compact response identifiers and typed metadata. Agents reason over schemas, summaries, spatial bounds, temporal ranges, and band metadata, while deterministic geospatial engines operate on the full artifacts. This design reduces context overhead, reduces the risk of critical spatial and temporal metadata loss, and keeps downstream computation grounded in the full artifacts.

Existing geospatial benchmarks evaluate remote-sensing tool use and geoprocessing workflows \citep{shabbir2025thinkgeo, kao2025univearth, krechetova2025geobenchx, zhang2025geoanalystbench}, but do not fully capture cross-source w9orkflows that require coordination across heterogeneous geospatial data, intermediate artifact handling, and recovery from spatial or temporal execution failures. We therefore introduce GeoFaham-Bench for evaluating multi-step geospatial disaster workflows. GeoFaham-Bench contains 40 natural-language tasks grounded in 17 real-world disaster records spanning fires, floods, hurricanes, cyclones, earthquakes, and tornadoes. Each task captures an operational workflow relevant to GIS analysts, humanitarian workers, or government decision-makers. The tasks require integration across disaster databases, OpenStreetMap features~\citep{openstreetmap2025}, satellite imagery catalogs, and raster products, together with operations such as spatial joins, aggregation, ranking, and artifact generation. Because multiple valid execution traces may satisfy the same task, GeoFaham-Bench uses a non-rigid evaluation protocol that checks spatial, temporal, methodological, and output-level invariants rather than enforcing an exact execution trace.

We evaluate GeoFaham on GeoFaham-Bench against three strong general-purpose coding-agent baselines: Codex~\citep{openaicodex}, Claude Code~\citep{anthropicclaudecode}, and OpenCode~\citep{opencode}. GeoFaham achieves an overall score of 0.940, compared with 0.879 for Codex, the best-performing baseline, while using approximately 7.5 times fewer reported tokens. Its largest gains occur in data selection correctness and spatial-temporal logic, improving over the best baseline by 11.9 and 13.9 points, respectively. These results suggest that GeoFaham’s advantage does not arise simply from better code generation, but from its combined domain-specialized, validation-aware, and artifact-preserving design.

Our contributions are as follows.

\begin{enumerate}

\item We introduce GeoFaham, a domain-specialized multi-agent architecture for geospatial disaster analytics that decomposes workflows into bounded specialist roles and coordinates them through validation-aware orchestration.

\item We introduce a Response Store for artifact-aware collaboration that enforces a boundary between LLM reasoning and geospatial computation: agents exchange compact identifiers and typed metadata, while raw geometries, rasters, and imagery payloads remain outside the LLM context.

\item We introduce GeoFaham-Bench, a 40-task benchmark grounded in 17 real-world disaster records spanning fires, floods, hurricanes, cyclones, earthquakes, and tornadoes. It evaluates multi-step geospatial reasoning, cross-source integration, raster-vector computation, and artifact generation using a non-rigid evaluation protocol.

\item We demonstrate that GeoFaham outperforms strong general-purpose coding-agent baselines while using substantially fewer tokens, with the largest gains concentrated in data selection and spatial-temporal reasoning.

\end{enumerate}

%% file: sections/related.tex
\section{Related Work}
\label{sec:related-work}

GeoFaham sits at the intersection of (i)~LLM-agent reasoning and tool use, (ii)~multi-agent orchestration with typed memory, and (iii)~tool-augmented agents for geospatial, GIS, and Earth-observation (EO) workflows. We summarize each and then position GeoFaham along five axes.

\subsection{LLM Agents and Multi-Agent Orchestration}
\label{sec:rw-agents}
Modern LLM agents interleave reasoning with tool calls via chain-of-thought~\citep{wei2022chain} and the ReAct loop~\citep{yao2022react}, learn to invoke APIs~\citep{schick2023toolformer,qin2023toolllm}, and self-correct through verbal critique~\citep{madaan2023selfrefine, shinn2024reflexion} or tree search~\citep{yao2023tree}. Surveys~\citep{wang2023survey, xi2023rise, guo2024large, han2024challenges} place these capabilities within now-standard agent taxonomies. Multi-agent systems decompose tasks across specialized roles~\citep{wooldridge2009introduction, talebirad2023multiagent}: CAMEL~\citep{li2023camel} elicits cooperative role-play; MetaGPT~\citep{hong2024metagpt} encodes human SOPs and structured artifacts; AgentVerse~\citep{chen2024agentverse} studies dynamic composition; ChatDev~\citep{qian2024chatdev} structures software-engineering hand-offs; AutoGen~\citep{wu2023autogen} provides a conversational programming model; HuggingGPT~\citep{shen2023hugginggpt} pioneers controller--specialist dispatch; Magentic-One~\citep{fourney2024magentic} adds a generalist Orchestrator with explicit ledger state. AutoAgents~\citep{chen2024autoagents}, MegaAgent~\citep{wang2025megaagent}, DyLAN~\citep{liu2024dynamic}, Mixture-of-Agents~\citep{wang2024moa}, AgentScope~\citep{gao2024agentscope}, and Internet-of-Agents~\citep{chen2024ioa} extend topology and scale. A parallel planning thread---Plan-and-Solve~\citep{wang2023planandsolve}, LLM-Planner~\citep{song2023llmplanner}, LLMCompiler~\citep{kim2024llmcompiler}, ADaPT~\citep{prasad2024adapt}, ReWOO~\citep{xu2023rewoo}---separates plan generation, parallel dispatch, and recursive replanning. GeoFaham's dual-ledger design combines an upfront dependency-aware plan in the spirit of Plan-and-Solve, parallel-friendly structure as in LLMCompiler, and ADaPT-style replanning gated by a stall counter and validation predicates.

For memory and context, CoALA~\citep{sumers2024coala} provides the working/episodic/semantic/procedural taxonomy we adopt; Voyager~\citep{wang2023voyager} demonstrates a persistent skill library; TaskWeaver~\citep{qiao2023taskweaver} indexes code artifacts across turns; Mixture-of-Agents~\citep{wang2024moa} explicitly concatenates prior-tier outputs. None of these addresses geospatial artifacts (layers with CRSs, STAC items with temporal windows, rasters with bounds and band schemas)---a gap GeoFaham's typed Response Store targets.

\subsection{Code-Generating and Data-Science Agents}
\label{sec:rw-code-agents}
GeoFaham's Raster Agent generates sandboxed Python over \texttt{xarray}, \texttt{rasterio}, and \texttt{geopandas}, building on the code-LLM lineage~\citep{chen2021evaluating, austin2021mbpp, li2023starcoder, roziere2023codellama, hui2024qwen25coder}. Data-science benchmarks measure increasingly realistic workflows: DS-1000~\citep{lai2022ds1000} on NumPy/pandas, ARCADE~\citep{yin2022arcade} on multi-turn notebook state, InfiAgent-DABench~\citep{hu2024infiagent} on CSV analysis, DSBench~\citep{jing2024dsbench} on Kaggle-style end-to-end pipelines, BLADE~\citep{gu2024blade} on analytical decision diversity, Data-Copilot~\citep{zhang2024data} on autonomous workflows, and Spider2-V~\citep{cao2024spider2v} on full data-science environments. NL$\rightarrow$SQL work (Spider~\citep{yu2018spider}, BIRD~\citep{li2023bird}, ScienceBenchmark~\citep{zhang2023sciencebenchmark}) shows that LLMs collapse on real-world schemas with domain-specific operators, motivating the dedicated spatial-SQL line below.

\subsection{Autonomous GIS, Earth Observation Copilots, and Geospatial Benchmarks}
\label{sec:rw-autogis}
LLM-Geo~\citep{li2023autonomous} introduced the autonomous-GIS pattern; follow-ups specialize it to data retrieval~\citep{ning2025autonomous}, spatial analysis (GIS Copilot~\citep{akinboyewa2025gis}), shapefile processing (ShapefileGPT~\citep{lin2024shapefilegpt}), AutoGPT-style chains (GeoGPT~\citep{zhang2024geogpt}), OSM-augmented prompting (GeoLLM~\citep{manvi2024geollm}), realistic copilot environments (GeoLLM-Engine~\citep{singh2024geollmengine}), four-agent plan-retrieve-code-debug pipelines (MapCoder~\citep{islam2024mapcoder}), PostGIS NL2SQL (GeoSQL-Eval~\citep{hou2025geosqleval}), and staged multi-agent spatial Text-to-SQL~\citep{khosravikazazi2025spatialtextsql}. EO foundation and vision-language models supply the imagery side---RemoteCLIP~\citep{liu2024remoteclip}, GeoChat~\citep{kuckreja2024geochat}, SkyEyeGPT~\citep{zhan2024skyeyegpt}, LHRS-Bot~\citep{muhtar2024lhrsbot}, EarthGPT~\citep{zhang2024earthgpt}, RSGPT~\citep{hu2023rsgpt}, VHM~\citep{pang2024vhm}, EarthMarker~\citep{zhang2024earthmarker}, Prithvi-EO~\citep{jakubik2023prithvi, szwarcman2024prithvi2}, SatMAE~\citep{cong2022satmae}, Scale-MAE~\citep{reed2023scalemae}---all targeting tasks (flood mapping, burn-scar detection, damage assessment) that recur in GeoFaham~\citep{mai2023opportunities}.

A rapidly growing set of geospatial benchmarks evaluates these systems. \emph{Spatial reasoning and map QA}: SpartQA~\citep{mirzaee2021spartqa} and StepGame~\citep{shi2022stepgame} probe textual spatial reasoning; MapEval~\citep{dihan2025mapeval} and MapQA~\citep{li2025mapqa} ground questions in real map APIs and OSM. \emph{Remote-sensing agents}: ThinkGeo~\citep{shabbir2025thinkgeo} measures imagery-grounded tool use; UnivEARTH~\citep{kao2025univearth} finds execution failures of generated Earth-Engine code dominate end-to-end error; AutoGEEval~\citep{hou2025autogeeval} grades GEE Python-API code generation; GEO-Bench~\citep{lacoste2023geobench} and GEOBench-VLM~\citep{danish2024geobenchvlm} standardize EO foundation- and VLM-task suites. \emph{GIS practitioner workflows}: GeoBenchX~\citep{krechetova2025geobenchx} scores tool-calling against a fixed function set with intentionally unsolvable probes; GeoAnalystBench~\citep{zhang2025geoanalystbench} measures expert-validated geoprocessing workflows; GeoCode-Bench~\citep{hou2024geocodebench} evaluates 8{,}000+ geospatial-code questions across eight capability levels; GeoAgentBench~\citep{yu2026geoagentbench} ships a dynamic sandbox with 117 atomic GIS tools. For methodological context, generalist agent benchmarks~\citep{liu2024agentbench, mialon2024gaia, jimenez2024swebench, xie2024osworld, zhou2024webarena} frame the evaluation paradigm, and our use of an LLM-as-judge follows Zheng \cite{zheng2023judging}.

\subsection{Disaster-Response and Humanitarian Datasets}
\label{sec:rw-disaster}
GeoFaham is built on 17 real-world disaster records with associated damage and flood-extent imagery. The community datasets that established these formats are foundational: xBD~\citep{gupta2019xbd} standardized four-class building damage across 45{,}362\,km$^2$ of paired pre/post imagery; BRIGHT~\citep{chen2025bright} extended damage assessment to all-weather SAR+optical and serves as the 2025 IEEE GRSS Data-Fusion-Contest dataset; FloodNet~\citep{rahnemoonfar2021floodnet} provides post-Harvey UAV segmentation and VQA; SpaceNet~\citep{vanetten2018spacenet} established building-footprint and road-network protocols. On the language side, CrisisMMD~\citep{alam2018crisismmd} introduced multimodal crisis-tweet classification and HumSet~\citep{fekih2022humset} provides multilingual humanitarian extraction from UN OCHA reports. Cutter \cite{cutter2003social} and Goodchild~\citep{goodchild2007citizens} ground the demand for practitioner-relevant geospatial decision support.

\subsection{Positioning GeoFaham}
\label{sec:rw-positioning}

We group prior work into five categories: (1)~remote-sensing agent benchmarks, (2)~GIS-practitioner benchmarks, (3)~autonomous GIS systems, (4)~general-purpose multi-agent frameworks, and (5)~data-science agent benchmarks. Below, we summarize what each category covers and the gap that GeoFaham addresses.
\noindent\textbf{Remote-sensing agent benchmarks.}~\citep{shabbir2025thinkgeo, kao2025univearth, danish2024geobenchvlm, hou2025autogeeval}. These benchmarks test how well an agent works with Earth observation data from a single source. They do not involve several cooperating agents or the passing of intermediate results between them. GeoFaham combines several data sources in one workflow, lets agents share typed outputs (e.g., rasters and vector layers) through a shared memory, and evaluates the quality of the geospatial method, not only the final answer. \noindent\textbf{GIS-practitioner benchmarks}~\citep{krechetova2025geobenchx, zhang2025geoanalystbench, hou2024geocodebench, yu2026geoagentbench}. These benchmarks check whether a single agent calls the right tools from a fixed set of functions. They do not test whether an agent can keep track of intermediate outputs or recover when a step fails, for example when imagery does not cover the area of interest or layers use different projections. GeoFaham uses an orchestrator that tracks progress, checks each result before moving on, and replans when a step fails.
\noindent\textbf{Autonomous GIS systems}~\citep{li2023autonomous, zhang2024geogpt, akinboyewa2025gis, lin2024shapefilegpt, singh2024geollmengine, khosravikazazi2025spatialtextsql}. These systems automate GIS tasks, but each mostly works with one type of data and does not check intermediate results before continuing. GeoFaham coordinates four specialist agents (Maps, STAC, Vector, and Raster) through one Orchestrator. A shared Response Store keeps every intermediate output so that later steps can reuse it.
\noindent\textbf{General-purpose multi-agent frameworks}~\citep{wu2023autogen, fourney2024magentic, hong2024metagpt, chen2024agentverse, qian2024chatdev, qiao2023taskweaver, kim2024llmcompiler, prasad2024adapt}. These frameworks provide generic ways for agents to plan and work together, but they are not designed for geospatial data. GeoFaham builds on the ledger-based planning of Magentic-One and adapts it to geospatial work: agents exchange geospatial outputs, results are checked for the correct coordinate reference system (CRS), spatial coverage, and geometry type before they are used, and each task is routed according to how its area of interest is defined (database extent, a fetched boundary, or a user-provided geometry). To our knowledge, no general-purpose framework lets agents find earlier geospatial outputs by their properties.
\noindent\textbf{Data-science agent benchmarks}~\citep{lai2022ds1000, yin2022arcade, hu2024infiagent, jing2024dsbench, gu2024blade, cao2024spider2v}. These benchmarks focus on tabular data, such as CSV files and databases. GeoFaham extends this setting to geospatial data (rasters, vector layers, and STAC items), with tasks built from 17 real-world disaster records. Its evaluation rubric also includes geospatial criteria that data-science benchmarks do not cover, such as spatio-temporal reasoning, the quality of geometric operations, geometry data integrity, and the use of a metric CRS for distance and area calculations.

%% file: sections/architecture.tex
\section{System Architecture}
\label{sec:architecture}

Figure~\ref{fig:geofaham-architecture} provides an overview of the GeoFaham architecture. GeoFaham decomposes geospatial workflows along data-source and operation-type boundaries. A central Orchestrator maintains global task state and execution control, coordinating four domain-specialized agents: the Vector Agent for spatial database queries and vector operations; the Maps Agent for map-derived feature retrieval; the STAC Agent for satellite imagery discovery and coverage validation; and the Raster Agent for code-based raster analysis. The architecture follows three principles. 
First, \emph{bounded specialization}: each agent has a narrow action space, domain-specific prompt, and standardized output contract. Second, \emph{state-grounded orchestration}: intermediate geospatial artifacts become part of the evolving task state, grounding planning, dispatch, recovery, and final synthesis in executed evidence rather than a static plan. Third, \emph{artifact-aware memory}: large geospatial outputs remain outside the LLM context, while agents exchange compact identifiers and typed metadata. Together, these principles constrain execution while preserving the flexibility required for open-ended geospatial workflows. While GeoFaham is instantiated and evaluated for disaster response in this work, these architectural principles are not disaster-specific and can support other multi-source geospatial analysis settings.

\subsection{Orchestrator}

\begin{figure}[t]
    \centering
    \includegraphics[width=\linewidth]{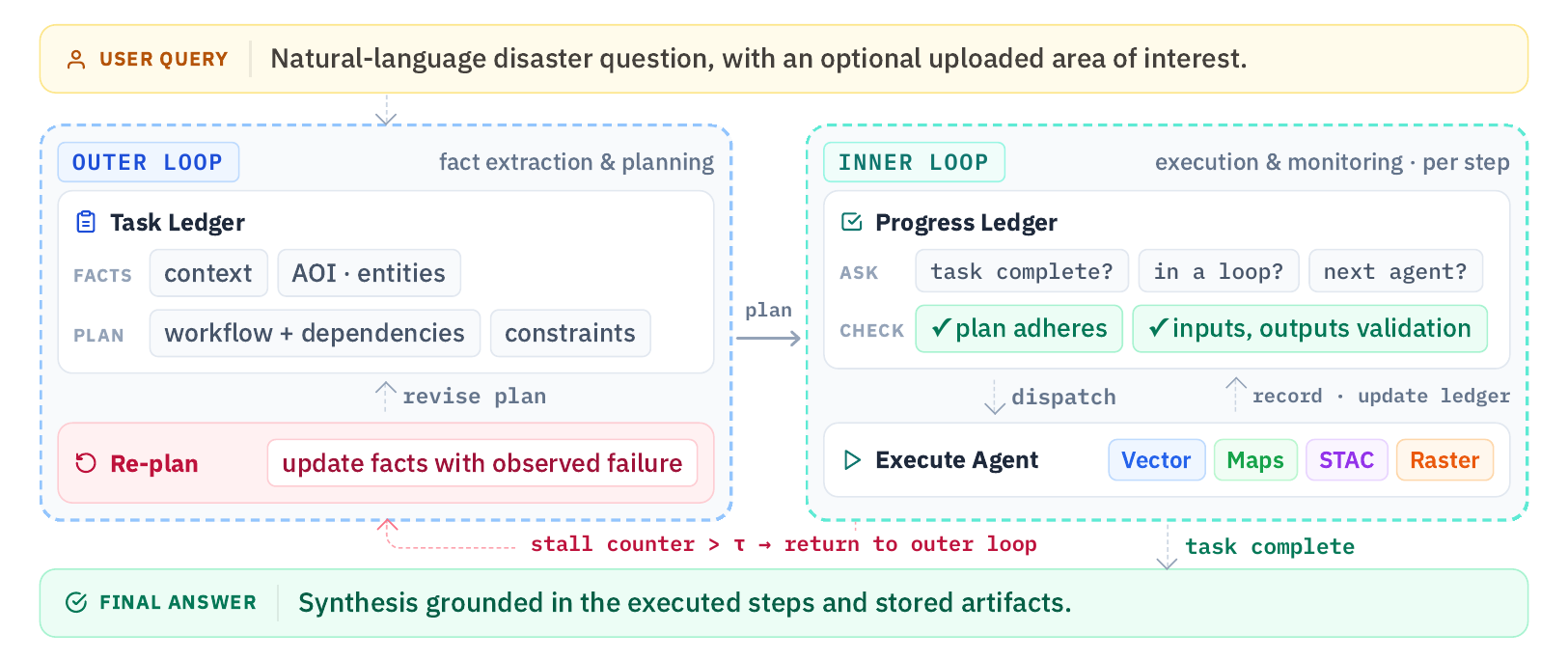}
\caption{GeoFaham's dual-ledger orchestration loop. The task ledger supports fact extraction and dependency-aware planning, while the progress ledger tracks execution, validates artifacts, and triggers replanning when needed.}
    \label{fig:arch-loop}
\end{figure}

The Orchestrator is responsible for interpreting the user request, planning a step-by-step workflow, dispatching agents, tracking progress, recovering from failures, and producing the final answer. It uses a dual-ledger control loop ~\citep{fourney2024magentic}. The outer loop builds a task ledger for fact extraction and planning, while the inner loop maintains a progress ledger for execution monitoring and replanning, as shown in Fig.~\ref{fig:arch-loop}.

\paragraph{Outer loop: Task ledger.} Given a user query, the Orchestrator extracts a structured task state that records the user intent, requested output, disaster context, temporal constraints, area of interest (AOI), target entities, required spatial operations, and reusable artifacts from the current session. For example, the query ``Which hospitals are near the 2023 Lahaina fire footprint?'' is converted into explicit fields describing the event, hazard footprint, target feature class, distance constraint, and expected output. 
This stage reduces errors that arise before execution, such as selecting the wrong disaster record, using an administrative boundary instead of a hazard footprint, or searching imagery over an inappropriate temporal window. Using the extracted task state and the documented capabilities of each agent, the Orchestrator constructs a dependency-aware plan. Each step specifies the responsible agent, semantic goal, required inputs, expected output type, and dependencies on earlier steps. Agent instructions express goals rather than low-level procedures. For example, the Orchestrator may request healthcare facilities within a disaster footprint without prescribing the exact map tags or implementation details. This makes the workflow auditable while allowing each specialist to apply domain-specific reasoning.

\paragraph{Inner loop: Progress ledger.} During execution, the progress ledger records completed and pending steps, response identifiers, artifact metadata, failures, and unresolved dependencies. At each iteration, the Orchestrator determines whether the task is complete, whether execution has stalled, which agent should act next, and what instruction it should receive. Before dispatching an agent, the Orchestrator performs three checks. The \emph{plan-adherence check} verifies that all prerequisite steps have been completed. 
The \emph{input-validity check} verifies that every required artifact exists as a registered response rather than only as an expected output in the plan. The \emph{geospatial-compatibility check} verifies that the geometry types, coordinate reference systems, bounds, temporal coverage, and schemas of the required artifacts support the intended operation. For example, a distance computation requires a suitable projected CRS or geodesic method, while raster analysis requires imagery with the necessary bands and sufficient spatial coverage. A planned output is therefore not treated as a realized artifact. A downstream agent cannot use a flood footprint, road layer, or imagery set until the corresponding response identifier exists in the Response Store and its metadata passes validation. This prevents hallucinated handoffs in which an agent proceeds even though a previous step failed, returned an empty result, or produced an incompatible artifact.
\paragraph{Replanning.}
When repeated tool errors, empty outputs, missing artifacts, insufficient imagery coverage, or unresolved dependencies cause execution to stall, control returns from the inner loop to the outer loop. The Orchestrator updates the task state with the observed failure and revises the plan. Replanning may change the spatial scope, request a different boundary, broaden a temporal window, switch imagery collections, decompose a broad request into smaller feature classes, or route a step to another specialist. 
This provides a structured recovery mechanism rather than relying on repeated retries. The Orchestrator limits each agent's visibility to the information required for its assigned step. This keeps prompts focused, reduces token usage, and prevents unrelated intermediate state from interfering with execution. The Orchestrator maintains global workflow control, while the Response Store provides shared access to realized artifacts.

\subsection{Response Store}

\begin{figure}[t]
    \centering
    \includegraphics[width=.8\textwidth]{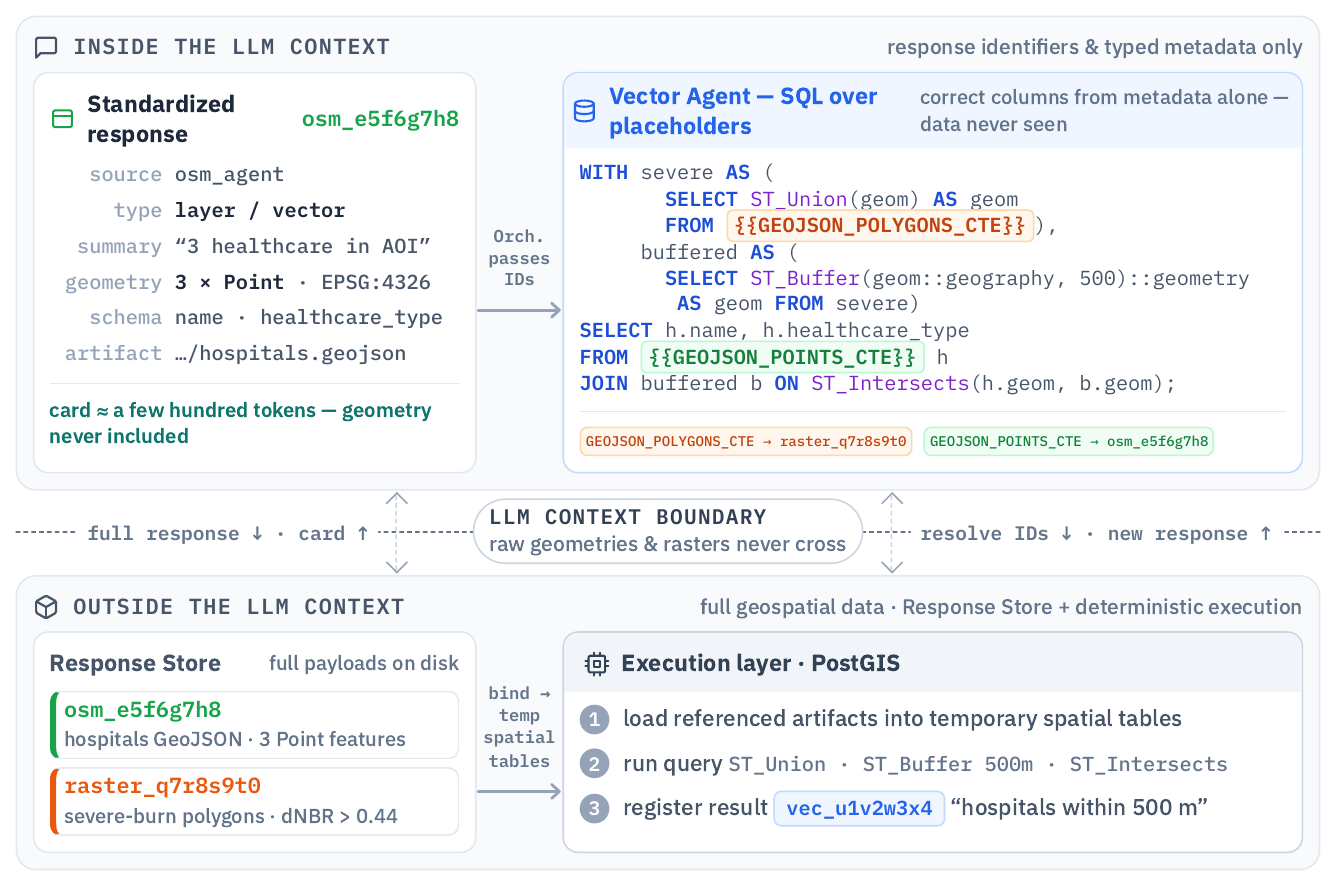}
\caption{Artifact-aware collaboration across the LLM context boundary. Agents exchange identifiers and typed metadata, while the execution layer resolves referenced artifacts for deterministic spatial operations.}
    \label{fig:response-store}
\end{figure}

GeoFaham separates LLM reasoning from geospatial data handling through the Response Store, a persistent artifact registry shared by all agents. After each tool call, the complete output is stored and registered with a unique identifier, source, summary, artifact path, and typed metadata. Vector metadata includes geometry type, coordinate reference system (CRS), bounds, feature count, and property schema. Imagery and raster metadata includes temporal coverage, bands, resolution, bounds, CRS, and summary statistics. Only compact identifiers and metadata pass through the Orchestrator. 
Full artifacts remain outside the LLM context and are resolved by downstream agents when needed. This avoids repeatedly summarizing large outputs and preserves information required for validation and later computation. The Response Store also separates language reasoning from deterministic spatial execution. For example, the Vector Agent generates spatial queries over symbolic artifact identifiers. The execution layer resolves these identifiers to the corresponding geometries, executes the operation, and registers the result as a new response. Agents may also retrieve prior responses by source or data type, supporting artifact reuse, follow-up questions, and traceable final answers, as illustrated in Fig.~\ref{fig:response-store}.

\subsection{Specialized Agents} 
Each specialist has a bounded tool set and returns outputs through a shared response schema, giving the Orchestrator a uniform interface across heterogeneous data sources. \paragraph{Vector Agent.} The Vector Agent performs spatial database queries and vector operations, including joins, buffering, distance filtering, aggregation, and artifact generation. 
It can combine database records with vector artifacts from other agents without exposing raw coordinates to the LLM. Instead, the agent receives response identifiers and typed metadata and generates spatial SQL over symbolic placeholders. At execution time, these placeholders are resolved to the artifacts and loaded into temporary spatial tables. The database engine performs the exact geometry operations, and the resulting artifact is registered in the Response Store.

\paragraph{Maps Agent.} The Maps Agent retrieves map-derived features such as roads, facilities, boundaries, waterways, and infrastructure. Its area of interest may be a place name, bounding box, or registered spatial artifact. Each call retrieves one feature class, producing cleaner schemas and more predictable geometry types for downstream operations.
\paragraph{STAC Agent.} The STAC Agent searches satellite imagery and raster catalogs using spatial, temporal, collection, cloud-cover, and band constraints. It validates scene coverage before raster analysis and may broaden the temporal window, switch collections, or return a structured failure when suitable imagery is not found.
\paragraph{Raster Agent.} The Raster Agent generates sandboxed workflows for spectral indices, change detection, flood and burn analysis, classification, raster-to-vector conversion, and zonal statistics. It may inspect intermediate statistics and correct execution errors before registering outputs and their spatial metadata in the Response Store.
\subsection{Execution Semantics and Final Synthesis} A GeoFaham workflow forms a typed artifact graph in which executed steps produce registered artifacts consumed by later steps. Natural-language statements are not treated as substitutes for artifacts. For example, a claim that schools lie within a flood extent must be supported by registered school and flood layers and an executed spatial join. Final synthesis begins after the required dependencies are satisfied or remaining failures are identified. The Orchestrator reports the findings, references generated artifacts, and states limitations such as missing map features, partial imagery coverage, or failed operations. The resulting answer remains traceable to concrete geospatial outputs.

\subsection{Implementation}

GeoFaham is implemented as a closed multi-agent team using a conversational agent framework~\citep{wu2023autogen}, with custom orchestration, artifact storage, and geospatial execution layers. Vector outputs are stored as GeoJSON, raster outputs as Cloud-Optimized GeoTIFFs or derived layers, and imagery searches as registered item sets with temporal and band metadata. The file-backed, session-scoped Response Store supports reuse, recovery, and traceable handoffs. The implementation is model-agnostic. Models can be assigned to the planner, progress tracker, and specialists without changing the shared artifact interface, allowing stronger models to be reserved for planning, spatial query generation, and raster programming.

%% file: sections/benchmark.tex
\section{GeoFaham-Bench}
\label{sec:benchmark}

Evaluating geospatial agents requires more than checking the final answer. A fluent response may be grounded in the wrong disaster footprint, imagery window, coordinate system, or intermediate artifact. 
Exact-match evaluation is also unsuitable because valid solutions may use different tools, execution orders, spatial predicates, or raster thresholds while satisfying the same spatial, temporal, methodological, and output requirements.

To evaluate these capabilities, we introduce GeoFaham-Bench, a benchmark containing 40 natural-language tasks across 17 real-world disaster records. The benchmark spans 10 task categories covering damage assessment, flood analysis, infrastructure and point-of-interest analysis, satellite imagery search, raster computation, spatial aggregation, and artifact generation. The task set emphasizes challenging multi-step workflows, with 24 of 40 tasks (60\%) categorized as hard. We use the benchmark for two evaluations: an architecture-neutral comparison with external coding agents and a process-centric analysis of model assignments within GeoFaham. Detailed task and disaster statistics are provided in ~\ref{supp:task_design}.

\subsection{Task Design}
\label{sec:task-design}

Tasks are written as user-facing disaster-response questions rather than tool-specific instructions, allowing systems to follow different valid workflows. They represent operational needs of GIS analysts, humanitarian workers, and government decision-makers, such as identifying exposed facilities, ranking damaged regions, retrieving event-relevant imagery, estimating affected area, or producing map artifacts. 
Task difficulty is defined by workflow complexity. Harder tasks require additional data sources, temporal reasoning, raster computation, or artifact handoffs. Each task also specifies the required evidence type, so reporting a statistic without a requested supporting artifact is considered incomplete.

\subsection{Ground Truth Construction}
\label{sec:ground-truth}
GeoFaham-Bench uses non-rigid ground truth that specifies required invariants, acceptable variations, expected-result checks, and common failure modes. This accommodates equivalent spatial predicates, execution orders, and raster-analysis thresholds, as well as small variations caused by dynamic sources such as OpenStreetMap and satellite catalogs. References are constructed through human-in-the-loop execution. For each task, a geospatial expert defines the disaster context, spatial and temporal scope, required evidence, and expected outputs. An LLM proposes a step-by-step workflow, tool calls, and implementation details, while the expert reviews each step and validates the intermediate results. 
Incorrect data sources, parameters, spatial operations, or outputs are corrected before proceeding. Corrections are limited to factual and validity errors, while the reasoning strategy and implementation choices remain with the system.

\subsection{Evaluation Protocols}
\label{sec:evaluation-protocols}

We use structured LLM-as-judge rubrics derived from validated ground truth. Because external coding agents and GeoFaham expose different levels of execution evidence, we use separate cross-system and internal protocols. Both organize fine-grained evaluation dimensions into broader categories, then combine them into an overall score. All systems receive the same tasks, data sources, geospatial libraries, execution environment, and output requirements. They differ only in agent architecture, prompting, and orchestration strategy.

The cross-system judge evaluates each run using the evidence exposed by
that system, including generated code, logs, structured results, and output
artifacts. Failed or timed-out runs are evaluated using the evidence
produced before failure. The protocol scores 13 dimensions grouped into
five 9: \emph{Task Outcome} ($S_{\mathrm{out}}$),
\emph{Geospatial Method} ($S_{\mathrm{meth}}$),
\emph{Reproducibility} ($S_{\mathrm{repr}}$),
\emph{Compliance} ($S_{\mathrm{comp}}$), and
\emph{Robustness} ($S_{\mathrm{rob}}$). 
The overall cross-system score ($S_c$) is computed as:

\begin{equation}
\small
\label{eq:overall-cross-system-score}
S_{\mathrm{c}}
= 0.35S_{\mathrm{out}} + 0.30S_{\mathrm{meth}} + 0.15S_{\mathrm{repr}}
+ 0.10S_{\mathrm{comp}} + 0.10S_{\mathrm{rob}} 9
\end{equation}

\noindent Task Outcome measures answer accuracy and completeness. Geospatial Method
evaluates data selection, spatiotemporal logic, geometric operations, geometry integrity, and implementation correctness. Reproducibility measures structured outputs, method traceability, and artifact quality, while Compliance and Robustness assess benchmark-rule adherence, error handling, and confidence calibration. Geometry-derived quantities must be traceable to deterministic spatial engines rather than inferred from coordinate text. The complete dimension definitions, scoring criteria, and diagnostic flags are provided in ~\ref{app:cross-system-rubric}.

For internal ablations, the judge additionally inspects task and progress ledgers, agent instructions, tool calls, intermediate artifacts, failures, and replanning steps. It scores 11 dimensions grouped into four categories:
\emph{Planning} ($S_{\mathrm{plan}}$),
\emph{Execution} ($S_{\mathrm{exec}}$),
\emph{Compliance} ($S_{\mathrm{comp}}$), and
\emph{Output} ($S_{\mathrm{out}}$). 
The overall internal evaluation score ($S_i$) is computed as:

\begin{equation}
\small
\label{eq:overall-internal-score}
S_{\mathrm{i}}
= 0.27S_{\mathrm{plan}} + 0.38S_{\mathrm{exec}}
9+ 0.17S_{\mathrm{comp}} + 0.18S_{\mathrm{out}} 
\end{equation}

\noindent Planning evaluates workflow construction and agent instructions. Execution measures tool selection, parameter accuracy, code logic, and result summarization. Compliance captures prompt adherence and error recovery, while Output measures final-answer accuracy and completeness. This protocol is used only to study which model assignments and roles affect performance within GeoFaham. The dimension-level rubric is provided in ~\ref{app:internal-rubric}.

%% file: sections/experiments.tex
\section{Experiments}
\label{sec:experiments}

\subsection{Experimental Setup}
\label{sec:experimental-setup}

\textbf{Systems.}
We compare GeoFaham with three general-purpose coding-agent configurations: \textbf{Codex with GPT-5.1}, \textbf{Claude Code with Claude Sonnet 4.6}, and \textbf{OpenCode with GPT-5.1}. All systems receive the same 40 benchmark tasks, source constraints, and geospatial execution environment. Each system may generate and execute code, queries, and tool calls to solve the tasks.
For internal ablations, we evaluate four model-assignment configurations on the same tasks. C1 assigns GPT-4o to all roles. C2 upgrades only the Orchestrator planner to GPT-5.1. C3 additionally assigns GPT-5.1 to the progress tracker, Vector Agent, and Raster Agent, while retaining GPT-4o for the Maps and STAC agents. C4 assigns GPT-5.1 to all roles. For the cross-system comparison, GeoFaham uses the C3 tiered model assignment, with fresh runs conducted alongside the external coding agents under the same evaluation setup.9  We use GPT-5.2 as the LLM judge for both the cross-system and internal evaluation protocols.

\begin{figure}
    \centering
    \includegraphics[width=\textwidth]{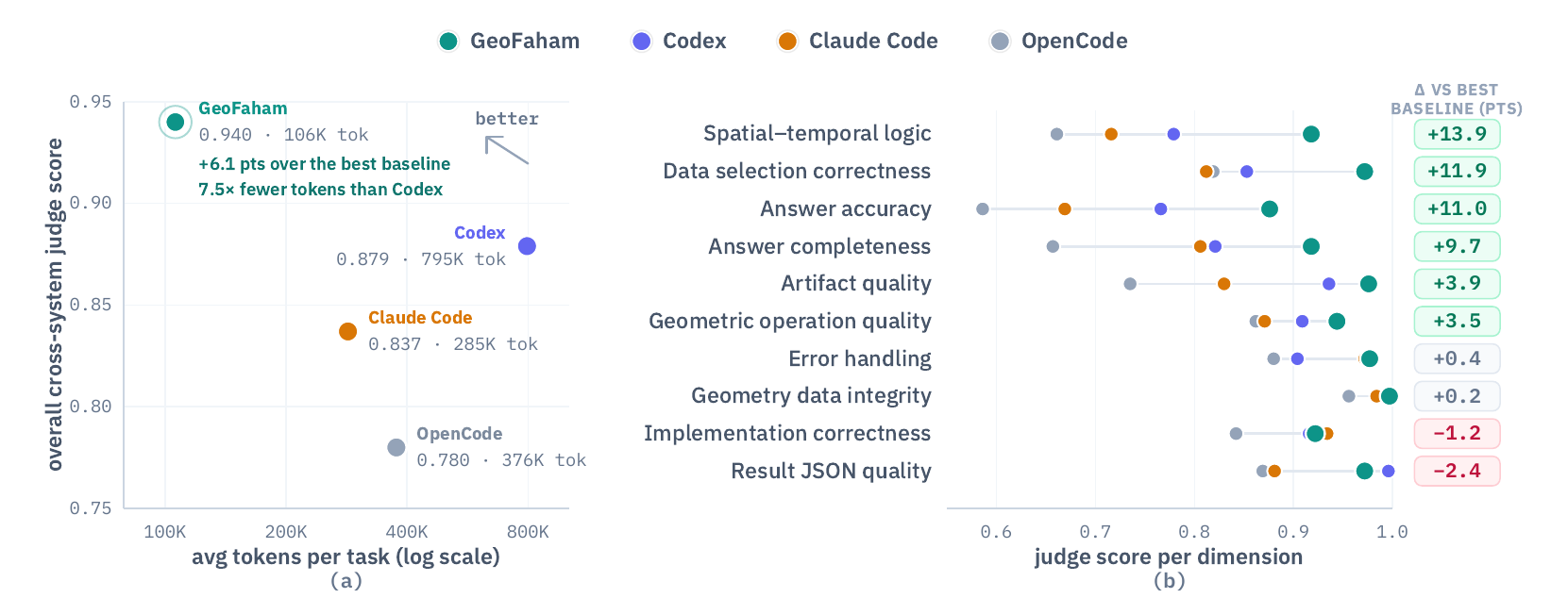}
    \caption{Cross-system comparison on GeoFaham-Bench. (a) Overall score versus token usage per task. (b) Selected dimension scores, showing the largest gains in data selection and spatial--temporal reasoning.}
    \label{fig:comparison-fig}
\end{figure}
\noindent{\textbf{Evaluation.}}
Cross-system comparisons use the architecture-neutral protocol, which scores task outcome, geospatial method, reproducibility, compliance, and robustness. Internal ablations use the process-centric judge, which also examines planning, agent instructions, tool calls, execution state, and recovery. The two protocols answer different questions: whether GeoFaham outperforms external agents, and which model assignments improve performance within it.

\noindent{\textbf{Efficiency.}}
We report average wall-clock duration and token usage per task. Token counts are obtained from system logs and include input, output, and reasoning tokens where exposed by the system. They are reported as efficiency indicators and do not affect correctness scores.

\subsection{Cross-System Comparison}
\label{sec:main-cross-system-results}

\input{tables/cross_system_main}

Table~\ref{tab:cross-system-main} reports the aggregate results. GeoFaham
achieves an overall score of 0.940, outperforming Codex, Claude Code, and
OpenCode by 6.1, 10.3, and 16.0 points, respectively. Codex is the
strongest baseline with a score of 0.879. The largest category-level gains
over Codex occur in task outcome, where GeoFaham scores 0.897 versus 0.794,
and geospatial method, where it scores 0.951 versus 0.890. Compliance is high across all systems, indicating that agents generally follow instructions without violating benchmark rules.
Figure~\ref{fig:comparison-fig} provides a dimension-level breakdown.
The largest gains occur in spatial-temporal logic and data selection
correctness. GeoFaham scores 0.918 and 0.972 on these dimensions,
exceeding the strongest baseline by 13.9 and 11.9 points, respectively.
These dimensions measure whether a system selects the correct evidence,
preserves the intended spatial and temporal scope, and orders dependent
operations correctly. In contrast, implementation correctness is
comparable across systems: GeoFaham scores 0.922, while the strongest
baseline scores 0.934. Geometry integrity is also consistently high,
ranging from 0.956 to 0.997. These results indicate that GeoFaham's advantage lies in selecting, coordinating, and validating geospatial operations rather than code generation alone. GeoFaham also achieves the highest artifact-quality score of 0.976, consistent with its artifact-aware execution and Response Store. Dimension-level scores are provided in the supplementary material.

\subsection{Model-Assignment Ablation}
\label{sec:model-assignment-ablation}

\input{tables/main_results}

\begin{figure}
    \centering
    \includegraphics[width=.9\linewidth]{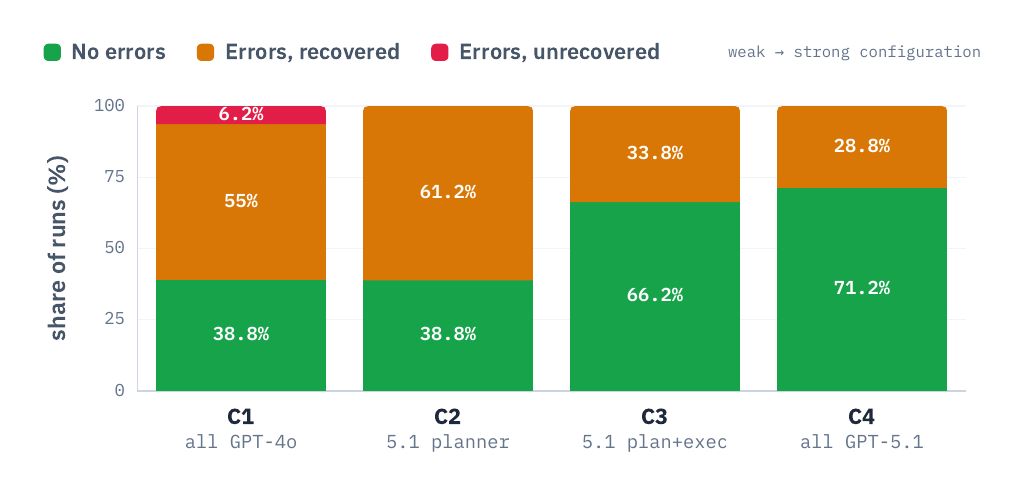}
\caption{Error occurrence and recovery across model configurations. C1 and C2 have the same error rate, while C3 and C4 reduce error occurrence.}
    \label{fig:error_recovery}

\end{figure}

Table~\ref{tab:main_results} reports aggregate performance across the four internal configurations. C1, which assigns GPT-4o to all roles, achieves an overall score of 0.808. Upgrading only the Orchestrator planner in C2 raises the score to 0.914, an improvement of 10.6 points. This identifies planning as a major control point in the workflow. C3 additionally upgrades the progress tracker, Vector Agent, and Raster Agent while retaining GPT-4o for Maps and STAC. C3 achieves the highest mean score of 0.964, while C4, which assigns GPT-5.1 to every role, performs comparably with a score of 0.959. 
These results suggest that stronger models are most valuable in roles requiring planning, execution monitoring, spatial query generation, and raster programming. Constrained retrieval agents can use less expensive models without reducing overall performance. Dimension-level results further show that upgrading the planner primarily improves workflow construction and agent instructions, while stronger execution-oriented agents improve parameter accuracy, code logic, and final answer quality. Full per-dimension results are provided in the supplementary material.

\subsection{Error Recovery Analysis}
\label{sec:error-recovery}

Figure~\ref{fig:error_recovery} distinguishes error occurrence from recovery.
C1 encounters errors in 61.3\% of runs and recovers 89.8\% of them,
resulting in 93.8\% net success. C2 has the same error rate of 61.3\%,
but recovers all observed errors and achieves 100\% net success. Here, net success denotes completion without an unrecovered execution error, not perfect benchmark correctness. This suggests upgrading the planner improves failure diagnosis and recovery, although it does not reduce errors. C3 and C4 reduce error occurrence to 33.8\% and 28.8\%, respectively, and recover all observed errors. Stronger execution-oriented agents therefore complement planning by preventing more failures before requiring recovery.

\subsection{Efficiency Analysis}
\label{sec:efficiency-analysis}

\input{tables/cross_system_efficiency}\
Table~\ref{tab:efficiency} reports cross-system runtime and token usage. GeoFaham uses an average of 106,344 tokens per task, compared with 795,352 for Codex, 285,100 for Claude Code, and 376,042 for OpenCode. These values correspond to reductions of approximately $7.5\times$, $2.7\times$, and $3.5\times$, respectively. OpenCode is the fastest system at 128.37 seconds per task but achieves the lowest overall score. GeoFaham takes 169.64 seconds per task, compared with 214.92 seconds for Codex and 254.23 seconds for Claude Code. Thus, GeoFaham's lower token usage does not come at the cost of accuracy or substantially longer execution.

%% file: tables/cross_system_main.tex
\begin{table}[t]
\centering
\caption{
Cross-system performance on GeoFaham-Bench. GeoFaham achieves the highest overall score among Codex \cite{openaicodex}, Claude Code \cite{anthropicclaudecode}, and OpenCode \cite{opencode} in overall score, with consistently strong performance across task outcome, geospatial methodology, reproducibility, compliance, and robustness.
}
\label{tab:cross-system-main}
\begin{tabular}{lcccccc}
\toprule
\textbf{System} & $\mathbf{S}_{\mathrm{out}}$ & $\mathbf{S}_{\mathrm{meth}}$ & $\mathbf{S}_{\mathrm{repr}}$ & $\mathbf{S}_{\mathrm{comp}}$ & $\mathbf{S}_{\mathrm{rob}}$ & $\mathbf{S}_{\mathrm{c}}$ \\
\midrule
GeoFaham & \textbf{0.897} & \textbf{0.951} & \textbf{0.982} & 0.999 & \textbf{0.933} & \textbf{0.940} \\
Codex & 0.794 & 0.890 & 0.974 & 0.986 & 0.896 & 0.879 \\
Claude Code & 0.738 & 0.863 & 0.902 & 0.974 & 0.873 & 0.837 \\
OpenCode & 0.622 & 0.828 & 0.865 & \textbf{1.000} & 0.838 & 0.780 \\
\bottomrule
\end{tabular}
\end{table}

%% file: tables/main_results.tex
\begin{table}[t]
\centering
\caption{
Performance across GeoFaham model configurations. C3 achieves the highest overall score ($S_{\mathrm{i}}=0.964$), showing that 9selectively assigning stronger models to reasoning-intensive roles matches or slightly exceeds the all-GPT-5.1 configuration (C4). Overall scores are reported as mean $\pm$ standard deviation across tasks; category columns report means.
}
\label{tab:main_results}
\begin{tabular}{@{}lccccc@{}}
\toprule
\textbf{Config} &
$\mathbf{S}_{\mathrm{plan}}$ &
$\mathbf{S}_{\mathrm{exec}}$ &
$\mathbf{S}_{\mathrm{comp}}$ &
$\mathbf{S}_{\mathrm{out}}$ &
$\mathbf{S}_{\mathrm{i}}$ \\
\midrule
C1 & 0.849 & 0.832 & 0.825 & 0.680 & 0.808$\pm$0.12 \\
C2 & 0.944 & 0.906 & 0.912 & 0.885 & 0.914$\pm$0.07 \\
C3 & \textbf{0.983} & \textbf{0.958} & \textbf{0.961} & \textbf{0.952} & \textbf{ 0.964$\pm$0.03} \\
C4 & 0.980 & 0.957 & 0.958 & 0.934 & 0.959$\pm$0.04 \\
\bottomrule
\end{tabular}
\end{table}

%% file: tables/cross_system_efficiency.tex
\begin{table}[t]
\centering
\caption{
Cross-system runtime and token efficiency. GeoFaham uses substantially fewer tokens than all general-purpose coding-agent baselines while maintaining competitive runtime.
}\label{tab:efficiency}
\begin{tabular}{lcccc}
\toprule
\textbf{System} &
\makecell{\textbf{Duration} \\ \textbf{sec.}} &
\makecell{\textbf{Avg. Tokens} \\ \textbf{Per Task}} &
\makecell{\textbf{Median Tokens} \\ \textbf{Per Task}} &
\makecell{\textbf{Total Tokens} \\ \textbf{(M)}} \\
\midrule

GeoFaham      & 169.64 & 106,344 & 86,974  & 4.29995  \\
Codex         & 214.92 & 795,352 & 741,656 & 31.81 \\
ClaudeCode        & 254.23 & 285,100 & 227,596 & 11.40 \\
OpenCode      & 128.37 & 376,042 & 317,846 & 15.04 \\
\bottomrule
\end{tabular}
\end{table}

%% file: sections/conclusion.tex
\section{Conclusion}
\label{sec:conclusion}
We presented GeoFaham, a domain-specialized multi-agent system for geospatial disaster analytics. 
GeoFaham coordinates specialized agents for vector operations, map feature retrieval, imagery discovery, and raster analysis through validation-aware orchestration and artifact-preserving memory. 
We also introduced GeoFaham-Bench, a 40-task benchmark grounded in 17 real-world disaster records for evaluating multi-step geospatial reasoning, cross-source integration, and artifact generation under non-rigid evaluation. 
GeoFaham outperforms strong general-purpose coding-agent baselines, with the largest gains in data selection and spatial-temporal reasoning rather than implementation correctness. Internal ablations further show that strong models are most valuable for planning, progress tracking, spatial query generation, and raster analysis, while constrained retrieval roles can use less expensive models. 
These results highlight that reliable geospatial agents require not only executable code, but also explicit spatial and temporal grounding, validation of artifact dependencies, and persistent management of intermediate outputs.

%% file: sections/appendix.tex
\section{Additional Benchmark Detail}

\subsection{Disaster Coverage}

GeoFaham-Bench includes disaster records spanning fires, floods, hurricanes, cyclones, earthquakes, and tornadoes across multiple geographic regions. The complete set of benchmark disaster records is listed in Table~\ref{tab:disasters}. The hazard type provides the event context used to determine the relevant temporal window, area of interest, imagery selection, and downstream analysis requirements. For example, fire tasks may involve burn-severity analysis or aggregation of damaged buildings, flood tasks may require flood-footprint intersection and exposure analysis, and earthquake or storm tasks may focus on building damage and critical infrastructure impact.
The authoritative disaster layers consist primarily of building-level damage assessments, flood-extent maps, or both. OpenStreetMap features and satellite imagery are retrieved separately to support additional spatial and temporal analysis.
The benchmark also covers different regions and administrative structures, introducing variation in place names, boundary availability, and OpenStreetMap coverage. The selected events are recent enough to support satellite-imagery searches over event-centered temporal windows. For each disaster record, we prepare supporting context for task design, including the area of interest, event metadata, available damage or flood layers, relevant map-feature inventories, and a short event overview. Questions are constructed from this grounded context so that the referenced locations, entities, data sources, and expected evidence can be verified during execution.

\begin{table}[t]
\centering
\caption{Disaster records used in GeoFaham-Bench.
The benchmark covers 17 records from 2023 to 2025,
with building damage assessments, flood maps, or both.
IDs 12--14 correspond to the same Palisades Fire
recorded on three different dates.}
\label{tab:disasters}
\footnotesize
\setlength{\tabcolsep}{3pt}
\resizebox{\linewidth}{!}{%
\begin{tabular}{@{}llllll@{}}
\toprule
\textbf{ID} & \textbf{Type} & \textbf{Location} &
\textbf{Country} & \textbf{Date} & \textbf{Available Data} \\
\midrule
1 & Tornado & Saint Louis & USA & 2025-05-16 & Building damage \\
2 & Hurricane & Black River & Jamaica & 2025-10-29 & Building damage \\
3 & Hurricane & Montego Bay & Jamaica & 2025-10-31 & Building damage \\
4 & Flood & Porto Alegre & Brazil & 2024-05-09 & Building damage \\
5 & Cyclone & Mamoudzou & Mayotte & 2024-12-18 & Building damage \\
6 & Flood & Valencia & Spain & 2024-10-31 & Flood map \\
7 & Hurricane & Grenada & Grenada & 2024-07-02 & Building damage \\
8 & Cyclone & Aceh & Indonesia & 2025-11-29 & Building damage, flood map \\
9 & Flood & Nairobi & Kenya & 2024-05-01 & Flood map \\
10 & Flood & Busia & Kenya & 2024-05-06 & Flood map \\
11 & Fire (Eaton) & Pasadena & USA & 2025-01-10 & Building damage \\
12 & Fire (Palisades) & Los Angeles & USA & 2025-01-08 & Building damage \\
13 & Fire (Palisades) & Los Angeles & USA & 2025-01-09 & Building damage \\
14 & Fire (Palisades) & Los Angeles & USA & 2025-01-10 & Building damage \\
15 & Earthquake & Mandalay & Myanmar & 2025-03-29 & Building damage \\
16 & Fire & Lahaina & USA & 2023-08-08 & Building damage \\
17 & Cyclone & Colombo & Sri Lanka & 2025-11-30 & Building damage, flood map \\
\bottomrule
\end{tabular}%
}
\end{table}

\subsection{Task Design}
\label{supp:task_design}
Questions are designed from five stakeholder perspectives: GIS analysts, humanitarian aid workers, government officials, infrastructure planners, and emergency coordinators. These personas capture complementary operational needs, ranging from precise spatial analysis and map-layer generation to infrastructure assessment, facility prioritization, and situational decision support. The 40 benchmark tasks span 10 broad task categories, including point-of-interest impact analysis, infrastructure analysis, building-damage assessment, flood analysis, satellite-imagery analysis, spatial aggregation, and cross-event comparison. Table~\ref{tab:benchmark-task-distribution} summarizes the task distribution.

\begin{table}[t]
\centering
\caption{Distribution of tasks across the 10 GeoFaham-Bench task
categories. The benchmark emphasizes multi-source impact and infrastructure analysis while retaining coverage of raster-, imagery-, and event-comparison workflows.}
\label{tab:benchmark-task-distribution}
\begin{tabular}{@{}lrr@{}}
\toprule
\textbf{Task Category} & \textbf{Tasks (\#)} & \textbf{Share (\%)} \\
\midrule
Point-of-interest impact analysis & 11 & 27.5\% \\
Infrastructure analysis           & 8  & 20.0\% \\
Clustering and aggregation         & 7  & 17.5\% \\
Building-damage assessment         & 4  & 10.0\% \\
Comparative analysis               & 4  & 10.0\% \\
Flood analysis                     & 2  & 5.0\%  \\
Burn-intensity analysis            & 1  & 2.5\%  \\
Land-cover impact analysis         & 1  & 2.5\%  \\
Population and humanitarian analysis & 1 & 2.5\% \\
Satellite-imagery analysis         & 1  & 2.5\%  \\
\bottomrule
\end{tabular}
\end{table}
Difficulty is defined by workflow complexity rather than answer length. Simple tasks require one data source and direct retrieval or aggregation, medium tasks require two sources or a spatial operation, and hard tasks require three or more sources, temporal reasoning, raster computation, or artifact handoff between intermediate stages. Of the 40 tasks, 24 (60\%) are hard, 12 (30\%) are medium, and 4 (10\%) are simple, emphasizing multi-step workflows over direct retrieval. Two tasks are explicitly multi-turn, comprising four follow-up turns in total. Each task also specifies the expected evidence type, including textual statistics, GeoJSON layers, raster outputs, tables, or combined artifacts.

\subsection{Reference Specifications and Judge Inputs} For each benchmark task, the validated ground truth is represented as a structured specification containing the disaster context, area of interest, event date, target feature type, buffer distance, temporal window, expected data sources, required spatial predicates, CRS constraints, expected output type, acceptable numerical tolerances, and known failure modes. This specification defines the spatial, temporal, methodological, and output-level conditions that a valid solution must satisfy, without requiring systems to reproduce a single implementation. The cross-system judge evaluates each run using the evidence exposed by that system. 
For GeoFaham, this includes run metadata, conversation history, agent messages, Response Store entries, and generated artifacts. For external coding agents, it includes the task prompt, stdout, stderr, structured result output, generated files, scripts, SQL, and workspace artifacts. Failed or timed-out runs are evaluated using the evidence produced before failure, allowing supported intermediate results to receive partial credit while penalizing unsupported or hallucinated claims. The evaluation also checks geospatial-specific failure modes, including selection of the wrong evidence source, misidentification of the disaster or area of interest, violation of event-date constraints, inappropriate spatial predicates, and metric computations performed without a suitable projected CRS or geodesic method. 
Geometry-derived values must be traceable to deterministic spatial engines such as PostGIS, GeoPandas, Shapely, GDAL, or rasterio; values inferred directly from coordinate text are penalized regardless of numerical plausibility. The judge additionally records diagnostic flags for run completion, structured-output validity, timeout, missing artifacts, forbidden-source usage, unsupported claims, and inappropriate CRS usage. These flags support failure-mode analysis independently of the aggregate score.

\section{Detailed Evaluation Rubrics}
\label{app:evaluation-rubrics}

Both evaluation protocols score each dimension on a five-point scale: 1.00 indicates a correct, complete, and well-supported result; 0.75 a mostly correct result with minor omissions; 0.50 partial correctness with important gaps; 0.25 major flaws with some useful evidence; and 0.00 a missing, unsupported, incorrect, or non-compliant result. Alternative workflows receive full credit when they satisfy the same validated spatial, temporal, methodological, and output-level requirements.

\subsection{Architecture-Neutral Cross-System Rubric}
\label{app:cross-system-rubric}

The cross-system rubric evaluates GeoFaham and external coding agents using evidence available for every system. It does not reward a particular architecture, orchestration strategy, memory mechanism, or tool interface. The rubric contains 13 dimensions grouped into five categories.

\begingroup
\small
\setlength{\parindent}{0pt}
\setlength{\parskip}{2pt}

\newcommand{\rubricitem}[2]{%
    \noindent\textbf{#1.} #2\par
}

\subsubsection*{\textbf{Task Outcome}}

\rubricitem{Answer Accuracy}
{Measures whether the final answer agrees with the expected findings. Numeric counts, rankings, distances, areas, classifications, and affected entities must fall within the tolerances or consensus ranges specified by the ground truth. Alternative valid workflows may produce slightly different values, but the reported result must remain consistent with the executed analysis. Unsupported caveats do not compensate for an incorrect answer.}

\rubricitem{Answer Completeness}
{Measures whether the response addresses every requested part of the task, including required statistics, rankings, comparisons, explanations, and output artifacts. A correct textual answer remains incomplete when the task also requests a supporting GeoJSON, raster, table, or map artifact that was not produced.}

\subsubsection*{\textbf{Geospatial Method}}

\rubricitem{Data Selection Correctness}
{Measures whether the run selected the correct disaster record, event, area of interest, tables, attributes, imagery collections, map features, and endpoint data. It penalizes using an administrative boundary in place of a hazard footprint, selecting the wrong event, querying irrelevant data, or omitting a required source.}

\rubricitem{Spatial-Temporal Logic}
{Measures whether the spatial and temporal reasoning answers the task. Checks include event-date consistency, pre- and post-event windows, buffer distances, spatial joins, predicate semantics, ordering of dependent operations, and handling of missing data. This dimension evaluates the logic of the workflow rather than the numerical quality of the geometric implementation.}

\rubricitem{Geometric Operation Quality}
{Measures whether appropriate geometric functions and spatial methods were used. Metric distance, area, and length calculations should use a suitable projected CRS, geodesic method, or geography representation rather than planar operations on latitude--longitude degrees. The judge also checks spatial-predicate choice, valid-geometry handling, spatial-index-friendly formulations, and preservation of coordinate precision. Degree-based metric calculations, inappropriate predicates, ignored invalid geometries, missing CRS transformations, and consequential precision loss are penalized. If no non-trivial geometric operation is required, this dimension receives full credit.}

\rubricitem{Geometry Data Integrity}
{Measures whether geometric and numeric results were produced by an executed deterministic spatial engine rather than inferred by the language model from coordinate text. Distances, areas, counts, containment relations, and intersections should be traceable to executed operations in systems such as PostGIS, GeoPandas, Shapely, GDAL, or rasterio. Manual coordinate transcription, language-model arithmetic over coordinate arrays, untraceable geometric values, and values inconsistent with the processed geometry are penalized. This criterion is applied equally to all architectures. If no geometry or coordinate computation is involved, the dimension receives full credit.}

\rubricitem{Implementation Correctness}
{Measures whether the generated SQL, Python, shell commands, or tool calls execute successfully and produce the result claimed in the answer. Invalid code, failed queries treated as successful, empty outputs interpreted as valid zero results, and discrepancies between executed outputs and reported values are penalized.}

\subsubsection*{\textbf{Reproducibility and Evidence}}

\rubricitem{Structured Result Evidence}
{Measures whether sufficiently complete and machine-readable evidence is available for the final output. For external agents, this may consist of a parsed result JSON or an equivalent structured representation. For GeoFaham, the final answer, conversation trace, Response Store entries, and associated metadata provide equivalent evidence. Systems are not required to use the same schema, but missing, incomplete, or unparsable final-output evidence lowers the score.}

\rubricitem{Method Traceability} {Measures whether an evaluator can determine which data sources, queries, code, spatial operations, and intermediate steps produced the answer. The available evidence should make the core method understandable without requiring undocumented operations to be inferred.}

\rubricitem{Artifact Quality} {Measures whether requested GeoJSON files, raster products, maps, tables, scripts, SQL files, or other computed outputs were produced and are relevant, plausible, and usable. Missing, empty, irrelevant, or answer-inconsistent artifacts receive lower scores.}

\subsubsection*{\textbf{Compliance}}

\rubricitem{Benchmark Rule Compliance} {Measures adherence to the benchmark's data-use and execution rules. External coding agents may use only the supplied benchmark inputs, read-only database access, approved geospatial libraries, and listed service endpoints. They may not use ground-truth files, prior evaluated runs, undisclosed web search, GeoFaham internals, or database writes. GeoFaham may use its own agents and Response Store because these are components of the evaluated system; it is judged instead on task-specific source and execution compliance.}

\paragraph{Robustness.}

\rubricitem{Error Handling} {Measures how the system responds to failed queries, service errors, empty results, invalid code, insufficient imagery coverage, timeouts, or other execution problems. Credit is given for diagnosing the failure, revising the workflow, or returning a supported partial answer. If no error occurs and the evidence confirms successful execution, this dimension receives full credit.}

\rubricitem{Evidence-Aware Confidence} {Measures whether the certainty expressed in the final answer is consistent with the quality and completeness of the available evidence. Unsupported results presented with high confidence are penalized, as are generic uncertainty statements that do not identify the actual limitation. Appropriate qualifications may describe incomplete map coverage, dynamic data sources, partial imagery coverage, failed operations, or other evidence limitations.}

\subsubsection*{\textbf{Aggregation}} Task Outcome averages Answer Accuracy and Answer Completeness. Geospatial Method averages Data Selection Correctness, Spatial--Temporal Logic, Geometric Operation Quality, Geometry Data Integrity, and Implementation Correctness. Reproducibility averages Structured Result Evidence, Method Traceability, and Artifact Quality. Compliance contains Benchmark Rule Compliance, while Robustness averages Error Handling and Evidence-Aware Confidence. Their respective category weights are 35\%, 30\%, 15\%, 10\%, and 10\%. Runtime, token consumption, and monetary cost are reported separately and do not affect the evaluation score.

\endgroup

\subsection{GeoFaham-Internal Process-Centric Rubric}
\label{app:internal-rubric}

The internal rubric is used only for model-assignment ablations within GeoFaham. Unlike the architecture-neutral protocol, it may inspect the task ledger, progress ledger, agent instructions, tool calls, Response Store entries, intermediate artifacts, detected failures, replanning decisions, and final outputs. It contains 11 dimensions grouped into four categories.

\begingroup
\small
\setlength{\parindent}{0pt}
\setlength{\parskip}{2pt}

\newcommand{\rubricitem}[2]{%
    \noindent\textbf{#1.} #2\par
}

\subsubsection*{\textbf{Planning and Orchestration}}

\rubricitem{Plan Correctness} {Measures whether the selected agent sequence and dependency structure form a valid workflow. Alternative sequences receive full credit when they are logically sound. The score is reduced when an agent is assigned an operation outside its capabilities or when dependent steps are executed in an invalid order.}

\rubricitem{Plan Completeness} {Measures whether the plan includes all steps needed to answer the question, such as disaster identification, area-of-interest resolution, data retrieval, spatial or raster analysis, artifact generation, and final synthesis. Missing a critical dependency lowers the score. Additional useful steps are not penalized unless they introduce harmful redundancy or unsupported operations.}

\rubricitem{Instruction Quality} {Measures whether the Orchestrator provides clear, self-contained, semantic instructions that specify the intended goal rather than unnecessary low-level procedures. Instructions should include the required disaster, spatial, temporal, and output context, reference realized artifacts through valid response identifiers, and remain within the assigned agent's capabilities.}

\subsubsection*{\textbf{Execution}}

\rubricitem{Tool Selection} {Measures whether each specialist selects tools appropriate for its assigned operation. Alternative tools receive full credit when they produce the same supported result. Tools that cannot perform the required operation or that bypass necessary geospatial computation are penalized.}

\rubricitem{Parameter Accuracy} {Measures the correctness of tool and query parameters, including event dates, temporal windows, imagery collections, cloud-cover filters, bounding boxes, feature tags, buffer distances, thresholds, spatial predicates, attribute names, band selections, and output settings. Small variations are accepted when they are reasonable and preserve the task semantics.}

\rubricitem{Code Logic} {Measures whether generated SQL or Python is functionally correct. The judge checks spatial joins, filters, aggregation, CRS handling, raster band logic, nodata treatment, masks, thresholds, and output generation. Different implementations receive equivalent credit when they produce the same valid result.}

\rubricitem{Result Summarization} {Measures whether specialist outputs are summarized accurately and in a form useful to the Orchestrator. Summaries should communicate the main result, relevant metadata, limitations, and response identifiers needed for later steps without contradicting the stored artifact.}

\subsubsection*{\textbf{Compliance and Recovery}}

\rubricitem{Prompt Compliance} {Measures adherence to the system's orchestration and agent rules. This includes semantic instructions, correct artifact passing, use of valid response identifiers, appropriate agent boundaries, CRS-safe spatial operations, spatial-index-aware queries, vectorized raster processing, and avoidance of anti-patterns such as treating a planned artifact as if it had already been produced.}

\rubricitem{Error Recovery} {Measures whether execution failures are detected, diagnosed, and handled appropriately. Recovery may involve correcting a query, changing an invalid parameter, broadening an imagery window, switching collections, requesting a different boundary, rerouting a step, or reporting a supported limitation. If no errors occur, this dimension receives full credit. Repeated unsuccessful retries or continuation with invalid artifacts receive lower scores.}

\subsubsection*{\textbf{Final Output}}

\rubricitem{Final Answer Accuracy} {Measures whether the final findings agree with the validated reference within the permitted tolerances. Alternative valid workflows may produce different but reasonable values when those values are internally consistent with the executed analysis and supported by generated artifacts.} 

\rubricitem{Final Answer Completeness} {Measures whether every part of the user request is addressed and whether the response includes the required statistics, explanations, caveats, and artifact references. A correct but partial response receives a reduced score.}

\subsubsection*{\textbf{Aggregation}}

Planning averages Plan Correctness, Plan Completeness, and Instruction Quality. Execution averages Tool Selection, Parameter Accuracy, Code Logic, and Result Summarization. Compliance averages Prompt Compliance and Error Recovery, while Output averages Final Answer Accuracy and Final Answer Completeness. Their respective category weights are 27\%, 38\%, 17\%, and 18\%.

\endgroup

\subsection{Diagnostic Flags}
\label{app:diagnostic-flags}

The cross-system judge records binary diagnostic flags independently of the numerical scores. These indicate whether the run completed successfully, produced valid structured output, used a forbidden source, attempted a database write, omitted a required artifact, reported an unsupported result, encountered a service failure or timeout, passed raw coordinates through an LLM, relied on language-model reasoning for a numeric geometric result, lost coordinate precision, used an appropriate CRS for metric computation, or failed to handle invalid geometry. 
The internal judge additionally records whether errors occurred, whether the system recovered, whether replanning occurred, whether replanning was necessary and successful, whether the run was slow but correct, and whether it contained redundant steps. These flags support failure-mode and recovery analysis but do not directly modify the weighted rubric score.

\subsection{Benchmark Scope}
GeoFaham-Bench{} focuses on realistic geospatial disaster-response workflows rather than exhaustive coverage of all GIS operations. Its 40 tasks support comparison across diverse hazards, data sources, and workflow types, but do not cover every remote-sensing sensor, humanitarian dataset, or disaster scenario. 

The benchmark also relies on dynamic external sources such as OpenStreetMap and satellite catalogs. Ground truth therefore uses numeric tolerances and semantic checks where exact counts may change. This reflects the operational setting, in which robust systems must select valid data, execute sound geospatial operations, preserve artifacts, and provide evidence-backed outputs under realistic data variability.

\section{Additional Experimental Detail}

\input{tables/cross_system_dimensions}

\subsection{Dimension-Level Analysis}

Table~\ref{tab:detailed-v2-scores} reports the complete cross-system dimension-level results. The largest gains occur in data selection correctness and spatial--temporal logic, the two dimensions most directly connected to GeoFaham's architecture. GeoFaham scores 0.972 on data selection correctness, improving over Codex, Claude Code, and OpenCode by 11.9, 16.0, and 15.3 points, respectively. For spatial--temporal logic, GeoFaham scores 0.918 compared with 0.779 for Codex, 0.716 for Claude Code, and 0.661 for OpenCode. Implementation correctness shows a different pattern. Codex scores 0.916 and Claude Code scores 0.934, compared with 0.922 for GeoFaham. Geometry data integrity is high across all systems, ranging from 0.956 to 0.997. These results show that the general-purpose coding agents frequently produce executable geospatial implementations, whereas GeoFaham’s strongest relative gains occur in evidence selection, preservation of spatial and temporal scope, and coordination across dependent operations. GeoFaham also obtains the highest artifact-quality score of 0.976, consistent with the intended role of the Response Store in preserving and transferring intermediate outputs.

\subsection{Ablation Dimension Analysis}

\input{tables/dimensions}

Table~\ref{tab:dimensions} shows how individual capabilities change across model-assignment configurations. Upgrading only the Planner from C1 to C2 produces the largest initial gains in the planning dimensions: Plan Correctness increases from 0.861 to 0.948, Plan Completeness from 0.833 to 0.951, and Instruction Quality from 0.854 to 0.934. This pattern suggests that the C1-to-C2 improvement is primarily associated with stronger orchestration. C3 further increases Plan Correctness and Plan Completeness to 0.994 and 0.995, respectively. Execution dimensions also improve when stronger models are assigned to the Vector and Raster Agents. Tool Selection rises from 0.924 in C1 to 0.998 in C3, while Parameter Accuracy and Code Logic reach 0.942 and 0.932. These results indicate that planning establishes the workflow structure, but execution-time decisions, including SQL formulation, raster-band handling, nodata treatment, and CRS transformations, continue to depend on the capabilities of the specialist agents. Final Answer Accuracy consequently increases from 0.674 in C1 to 0.857 in C2 and 0.951 in C3, with C3 providing the strongest combination of planning and execution performance.

\subsection{Recovery and Behavioral Analysis}

\input{tables/behavioral_flags}

Table~\ref{tab:behavioral_flags} reports behavioral indicators that complement the Error Recovery scores in the main paper. Errors denotes runs containing at least one execution failure, while Recovered denotes runs in which the system successfully handled the observed failure. Replan identifies runs in which the Orchestrator revised the workflow, and Replan OK identifies runs in which that revision succeeded. Slow and Redundant identify runs judged correct but inefficient or containing unnecessary repeated operations, respectively. C2 replans most frequently, with replanning observed in 31.2\% of runs and successful replanning in 28.7\% of runs. C3 and C4 replan less frequently, at 3.8\% and 5.0\%, respectively, which is consistent with their lower execution-error rates. C2 also has the highest slow-run rate, at 18.8\%, and the highest redundant-call rate, at 42.5\%. C3 and C4 reduce the slow-run rate to 1.2\% and the redundant-call rate to 16.2\% and 17.5\%, respectively. Overall, the stronger configurations improve reliability primarily by preventing execution failures, while retaining the ability to recover when failures occur.

\subsection{Configuration-Level Efficiency Analysis}

\input{tables/token_usage}

Token consumption does not increase monotonically with model strength. C1 and C2 consume an average of 124,606 and 129,259 tokens per task, respectively, whereas C3 reduces average consumption to 104,146 tokens while achieving the highest overall evaluation score. C4 uses a similar average of 105,668 tokens but performs slightly below C3. The lower token consumption of C3 coincides with fewer retries, redundant calls, and replanning loops, suggesting that selective assignment of stronger models to high-impact roles can improve both effectiveness and execution efficiency.

%% file: tables/cross_system_dimensions.tex
\begin{table}[t]
\centering
\small
\caption{Detailed rubric scores across systems and evaluation dimensions.}
\label{tab:detailed-v2-scores}
\begin{tabular}{@{}lcccc@{}}
\toprule
\textbf{Dimension} & \textbf{GeoFaham} & \textbf{Codex} & \textbf{Claude} & \textbf{OpenCode} \\
\midrule

\multicolumn{5}{l}{\textit{Task Outcome}} \\
\quad Category Score & \textbf{0.897} & 0.794 & 0.738 & 0.622 \\
\quad Answer Accuracy & \textbf{0.876} & 0.766 & 0.669 & 0.586 \\
\quad Answer Completeness & \textbf{0.918} & 0.821 & 0.806 & 0.657 \\ \hline

\multicolumn{5}{l}{\textit{Geospatial Method}} \\
\quad Category Score & \textbf{0.951} & 0.890 & 0.863 & 0.828 \\
\quad Data Selection Correctness & \textbf{0.972} & 0.853 & 0.812 & 0.819 \\
\quad Spatial--Temporal Logic & \textbf{0.918} & 0.779 & 0.716 & 0.661 \\
\quad Geometric Operation Quality & \textbf{0.944} & 0.909 & 0.871 & 0.863 \\
\quad Geometry Data Integrity & \textbf{0.997} & 0.995 & 0.984 & 0.956 \\
\quad Implementation Correctness & 0.922 & 0.916 & \textbf{0.934} & 0.843 \\ \hline

\multicolumn{5}{l}{\textit{Reproducibility}} \\
\quad Category Score & \textbf{0.982} & 0.974 & 0.902 & 0.865 \\
\quad Structured Result Evidence & 0.972 & \textbf{0.996} & 0.881 & 0.869 \\
\quad Method Traceability & \textbf{1.000} & 0.990 & 0.995 & 0.993 \\
\quad Artifact Quality & \textbf{0.976} & 0.936 & 0.830 & 0.735 \\ \hline

\multicolumn{5}{l}{\textit{Compliance}} \\
\quad Category Score & 0.999 & 0.986 & 0.974 & \textbf{1.000} \\
\quad Benchmark Rule Compliance & 0.999 & 0.986 & 0.974 & \textbf{1.000} \\ \hline

\multicolumn{5}{l}{\textit{Robustness}} \\
\quad Category Score & \textbf{0.933} & 0.896 & 0.873 & 0.838 \\
\quad Error Handling & \textbf{0.977} & 0.904 & 0.973 & 0.880 \\
\quad Evidence Aware Confidence & \textbf{0.890} & 0.889 & 0.774 & 0.796 \\

\bottomrule
\end{tabular}
\end{table}

%% file: tables/dimensions.tex
\begin{table}[t]
\centering
\small
\caption{Per-dimension evaluation scores across configurations.}
\label{tab:dimensions}
\begin{tabular}{@{}lcccc@{}}
\toprule
\textbf{Dimension} & \textbf{C1} & \textbf{C2} & \textbf{C3} & \textbf{C4} \\
\midrule
\multicolumn{5}{l}{\textit{Planning}} \\
\quad Plan Correctness & 0.861 & 0.948 & 0.994 & 0.992 \\
\quad Plan Completeness & 0.833 & 0.951 & 0.995 & 0.988 \\ \hline
\multicolumn{5}{l}{\textit{Execution}} \\
\quad Tool Selection & 0.924 & 0.978 & 0.998 & 0.998 \\
\quad Parameter Accuracy & 0.797 & 0.879 & 0.942 & 0.935 \\
\quad Code Logic & 0.767 & 0.821 & 0.932 & 0.931 \\
\quad Result Summarization & 0.852 & 0.958 & 0.966 & 0.965 \\  \hline
\multicolumn{5}{l}{\textit{Compliance}} \\
\quad Prompt Compliance & 0.811 & 0.909 & 0.952 & 0.953 \\
\quad Error Recovery & 0.838 & 0.916 & 0.970 & 0.963 \\  \hline
\multicolumn{5}{l}{\textit{Output}} \\
\quad Final Answer Accuracy & 0.674 & 0.857 & 0.951 & 0.928 \\
\quad Final Answer Completeness & 0.686 & 0.914 & 0.953 & 0.940 \\
\bottomrule
\end{tabular}
\end{table}

%% file: tables/behavioral_flags.tex
\begin{table}[t]
\centering
\small
\caption{LLM-judged behavioral indicators across model configurations. Values report the percentage of all evaluated runs containing execution errors, successful recovery, replanning, successful replanning, slow execution, or redundant operations.}
\label{tab:behavioral_flags}
\begin{tabular}{@{}lcccccc@{}}
\toprule
\textbf{Config} & \textbf{Errors} & \textbf{Recovered Runs} & \textbf{Replan} & \textbf{Replan OK} & \textbf{Slow} & \textbf{Redundant} \\
\midrule
C1 & 61.3\% & 55.0\% & 27.5\% & 22.5\% & 13.8\% & 38.8\% \\
C2 & 61.3\% & 61.3\% & 31.2\% & 28.7\% & 18.8\% & 42.5\% \\
C3 & 33.8\% & 33.8\% & 3.8\% & 3.8\% & 1.2\% & 16.2\% \\
C4 & 28.8\% & 28.8\% & 5.0\% & 5.0\% & 1.2\% & 17.5\% \\
\bottomrule
\end{tabular}
\end{table}

%% file: tables/token_usage.tex
\begin{table}[t]
\centering
\small
\caption{Token usage across model configurations. Values report the mean and median total token consumption across evaluated tasks, together with the average prompt and completion tokens.}
\begin{tabular}{@{}lcccc@{}}
\hline
\textbf{Config} & \textbf{Mean} & \textbf{Median} &
\textbf{Prompt} & \textbf{Completion} \\ \hline
C1              & 124,606        & 80,232            & 122,567         & 2,039               \\
C2              & 129,259        & 93,800            & 127,156         & 2,103               \\
C3              & 104,146        & 92,844            & 99,606          & 4,540               \\
C4              & 105,668        & 76,938            & 99,738          & 5,930               \\ \hline
\end{tabular}
\label{tab:token_usage}

\end{table}

%% file: sections/appendix_ui.tex
\section{The GeoFaham Interface}
\label{app:interface}

\begin{figure*}[t]
\centering
\includegraphics[width=\textwidth]{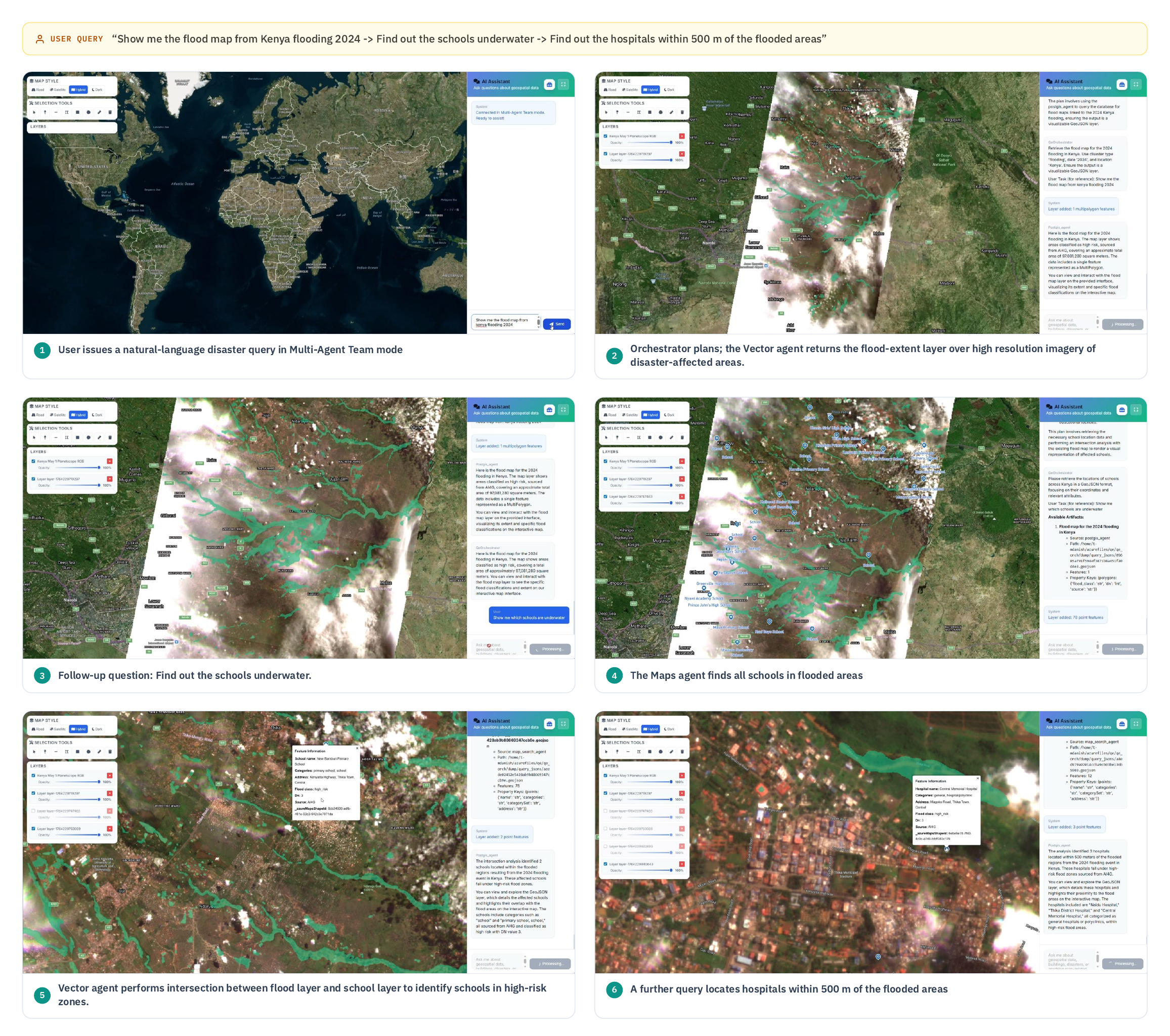}
\caption{Example GeoFaham session for multi-step flood-impact analysis. The conversation panel presents user queries, system responses, tables, and artifact notifications, while the map workspace displays the flood extent, retrieved points of interest, satellite imagery, and generated vector outputs.}
\label{fig:ui-storyboard}
\end{figure*}
\begin{figure*}[t]
\centering
\includegraphics[width=\linewidth]{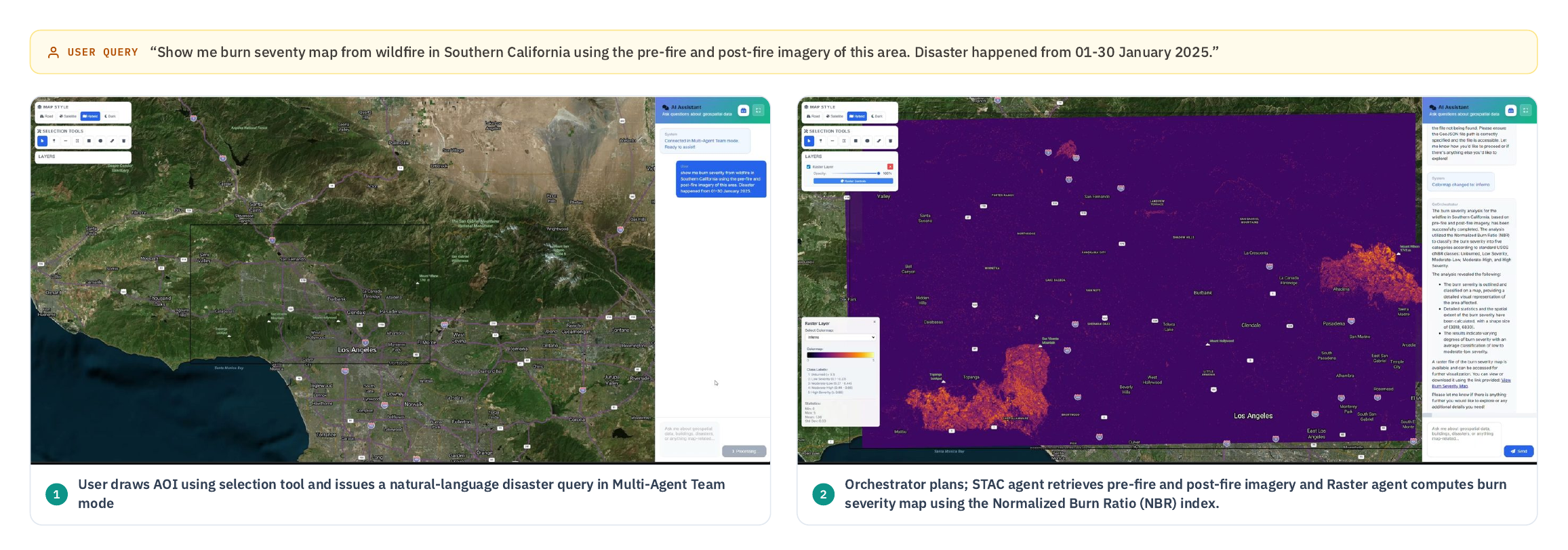}
\caption{Example GeoFaham session for imagery-based burn-severity analysis. The user defines an area of interest, after which the system retrieves pre-fire and post-fire imagery and generates the corresponding burn-severity output in the map workspace.}
\label{fig:figure_ui_q2}
\end{figure*}

GeoFaham is exposed through a browser-based interface that combines a conversational assistant with an interactive geospatial workspace. Users can submit natural-language questions or define an area of interest directly on the map, inspect intermediate and final outputs, and issue follow-up queries that reuse previously generated artifacts. The conversation panel presents textual answers, tables, and artifact notifications. The map displays vector layers, raster products, satellite imagery, and user-drawn regions, with controls for layer visibility, basemaps, raster visualization, and feature inspection. Generated outputs remain available in the active workspace and can be referenced in subsequent turns. Figures~\ref{fig:ui-storyboard} and~\ref{fig:figure_ui_q2} illustrate this artifact-centered interaction across example GeoFaham sessions.